 \documentstyle[epsf,prd,aps,floats]{revtex} 
  
\newcommand{\be}{\begin{equation}}   
\newcommand{\ee}{\end{equation}}   
\newcommand{\bea}{\begin{eqnarray}}   
\newcommand{\eea}{\end{eqnarray}}

\begin{document}    
\twocolumn[\hsize\textwidth\columnwidth\hsize\csname@twocolumnfalse\endcsname    
\author{Jo\~ao Magueijo}   
\date{\today} 
\address{Theoretical Physics, The Blackett Laboratory, 
Imperial College, Prince Consort Road, London, SW7 2BZ, U.K.} 
\title{Covariant and locally Lorentz-invariant varying speed of light   
theories}  
 
\maketitle    
\begin{abstract} 
We propose definitions for covariance and local Lorentz invariance  
applicable when the speed of light $c$ is allowed to vary. They have  
the merit of retaining only those aspects of the usual definitions which are  
invariant under unit transformations, and which can therefore  
legitimately represent
the outcome of an experiment. We then discuss some possibilities  
for invariant actions governing the dynamics of such theories.  
We consider first the classical action for matter fields and  
the effects of a changing $c$ upon quantization. We discover a 
peculiar form of quantum particle creation due to a varying $c$. 
We then study actions governing the dynamics of gravitation and  
the speed of light. We find the free, empty-space, no-gravity solution,  
to be interpreted as the counterpart of Minkowksi space-time,  and   
highlight its similarities with Fock-Lorentz space-time. We also  
find flat-space string-type solutions, in which near the  
string core $c$ is much higher. We label them fast-tracks and
compare them with gravitational wormholes. We finally discuss general 
features of cosmological and black hole solutions, and digress on 
the meaning of singularities in these theories.  
\end{abstract}   
\pacs{PACS Numbers: *** } 
 ]

\renewcommand{\thefootnote}{\arabic{footnote}} 
\setcounter{footnote}{0}

\section{Introduction}    
The varying speed of light (VSL) theory provides an elegant   
solution to the cosmological problems - the horizon, flatness, and Lambda  
problems of Big-Bang cosmology. The theory has appeared in several   
guises 
(\cite{mof1,am,b,bm1,bm2,bm3,bm4,mof2,cly1,cly2,cly3,drum,av1,av2,harko,kir,alex,bass})), but in the formulation proposed    
by Albrecht and Magueijo \cite{am} (see also 
\cite{b,bm1,bm2,bm3,bm4,av1,av2}) one finds  
the most direct mechanism for converting the Einstein   
deSitter model into a cosmological attractor. Unfortunately the   
foundations of such a theory are far from solid. Covariance and local  
Lorentz invariance are explicitly broken, and are not replaced by similar  
far-reaching principles. The difficulty in applying the theory to   
situations other than cosmology (eg. black holes) stems directly from   
this deficiency.  
  
This paper is an attempt to remedy this shortcoming. This may be achieved  
in various different ways, some of which inevitably rather  
radical. We   
note that nothing prevents the construction of a theory satisfying  
the principle of relativity, while still allowing for space-time 
variations in $c$. Such a theory would in general not be  
Lorentz invariant, but it could still   
be relativistic. Indeed, Lorentz invariance follows from two independent  
postulates: the principle of relativity and the principle of constancy 
of the   
speed of light. Dropping the latter while keeping the former leads to a   
new invariance, known as Fock-Lorentz symmetry \cite{fock,man,step}. 
This invariance does  
not distinguish between inertial frames (and therefore satisfies the  
principle of relativity) but it allows for a   
varying $c$; indeed it allows for a non-invariant $c$. 
  
A possible approach is therefore to set up a  
theory of gravitation based upon a gauged Fock-Lorentz  
symmetry. However we note that such an enterprise 
accommodates more than is required by VSL theories:   
it allows  the speed of light at a given point to depend  
on the observer's speed. Also the speed of light in the Fock-Lorentz space  
is anisotropic. Clearly, certain aspects of the second postulate of  
Einstein's relativity theory may be kept in the simplest VSL theories,  
namely that the speed of light at a given point be independent of  
its color, direction, or the speeds of either emitter or observer.   
  
In this paper we shall be as conservative as possible and preserve   
all aspects of the second postulate of special relativity consistent  
with allowing space-time variations of $c$. In Section~\ref{cov} we show  
that such a reformulation gleans from the second postulate of relativity 
all that is operationally meaningful, in the sense that  
the aspects of the second postulate 
which we preserve are exactly those which 
can be the outcome of experiment (such 
as the Michelson-Morley experiment). The constancy of $c$ in space-time, 
on the other hand, 
amounts to nothing more than a definition of a system of units.  
In Section~\ref{cov} and \ref{cov1} 
we show that such a theory is {\it locally} Lorentz invariant  
and generally covariant, subject to a minimal generalization  
of these concepts. In Section~\ref{cov2} we summarise the overall 
structure of such theories, and the basic reasons for adopting it. 
  
We then discuss Lagrangians governing the dynamics of these   
theories. The main practical drawback of explicit lack of covariance  
is that it makes an action principle formulation rather awkward 
(see \cite{bm1,bass}).  
The VSL theories proposed in this paper, on the contrary,  
are easily amenable to an action principle formulation. However 
we shall try to borrow some features from earlier models, such as 
lack of energy conservation.  
 
In Section~\ref{matter} we first consider the matter action. We show how  
it is always possible to define the matter Lagrangian so that $c$ does not  
appear explicitly. Such a principle fixes a large number of scaling laws 
for other ``constants'' as a function of $c$. It also leads to simpler 
dynamical equations for $c$.  
 
Two constants are left undetermined by these considerations: Planck's  
constant $\hbar$ and Boltzmann's constant $k_B$. These cannot be determined 
by classical dynamics, and scaling laws $\hbar(c)$ and $k_B(c)$ should 
be postulated. In Section~\ref{quanta} we consider the implications 
of various $\hbar(c)$ for quantization. We identify situations 
in which a varying speed of light leads to quantum particle creation. 
 
Then in Section~\ref{grav} we consider Lagrangians for gravitational 
dynamics (we include the dynamics of $c$ into this discussion - as 
the field $c$ can be seen as an extra gravitational field). We identify 
the actions which lead to nothing but a change of units in a standard 
Brans-Dicke theory; all other actions are intrinsically different theories. 
 
The rest of the paper is devoted to the simplest applications of these
theories. In Section \ref{empty} we discuss empty space solutions. We find
a variation in $c$ and a global space-time which is very similar to
those found in Fock-Lorentz space. We also show how Fock-Lorentz space 
is nothing but a change of units applied to Minkowski space-time. 
However $t=\infty$ is brought to a finite time in the varying $c$ 
representation. We show that the space is actually extendable beyond 
this finite time; into what in the fixed $c$ representation would be a 
trans-eternal region. 

Another flat-space solution to our theory is a soliton string, close
to which the speed of light is much larger. We label it a fast-track.
A spaceship moving along
a fast track could move at non-relativistic speeds, without a twin
paradox effect, and still cover enormous intergalactic distances.
These solutions are not dissimilar to gravitational wormholes;
and indeed they are mapped into wormhole-like structures in fixed
$c$ units.

We finally discuss general features of cosmological solutions and
black holes in these theories. These will be developed further
in two publications 
currently under  preparation \cite{vcb,vcc}. Concerning
black holes the main novelty is that
for some regions of the couplings the speed of light may go to
zero at the horizon. This effectively prevents any observer from
entering the horizon, and its interior should therefore be excised 
from the manifold. We relabel this boundary an ``edge'', and 
comment on the implication of this effect for a generalized cosmic
censorship principle.

\section{Generalized Lorentz invariance }  
\label{cov}  
From an operational point of view all laws of physics should be   
invariant under global and local changes of units \cite{dick1,bek1,am}.  
Indeed measurements are always ratios to standard units, and therefore  
represent essentially dimensionless quantities. Physics should  
therefore be dimensionless or unit-invariant.   
However, this far-reaching principle is rarely incorporated into    
theoretical constructions, because a concrete choice of units usually  
simplifies the statement of laws. While this practical consideration  
should be recognized, it is important to realize  
that some theoretical constructions are tautological, and amount to  
nothing more than the specification of  a system  of units.  
  
An example is the second postulate of special  
relativity: the constancy of $c$. Clearly the postulate is invariant  
under unit transformations when it states that light of different colors  
travels at the same speed - as it makes a statement about the ratio of two  
speeds at a given point, which is a dimensionless quantity.   
The postulate is also unit-independent when it incorporates the result   
of the Michelson-Morley experiment: light emitted by sources moving at   
different speeds travels at the same speed. Again it makes use of ratios   
of speeds: the ratio of the sources' speeds, and the ratios of the   
different light rays' speeds. However the postulate looses its  
meaning when it refers to light speed at different points,  
or even to the speed of light moving in different directions  
at a given point.   
  
Hence Lorentz invariance in its usual definition   
is not a unit-independent concept, and indeed  
relativity is not a unit-independent construction. For instance relativity  
is not conformally invariant (a conformal transformation  
being just a particular type of unit transformation). A unit-independent  
definition of Lorentz invariance may be inferred by taking a Lorentz   
invariant theory and subjecting it to the most general unit transformation.  
The resulting theory retains the unit-invariant aspects of the   
second postulate, and clearly $c$ may now be anisotropic and vary in   
space-time. Under such circumstances what is the structure  
which represents Lorentz invariance?

For simplicity we specialize to 
changes of units which only affect the local value  
of the modulus of $c$. We redefine units of time and space  
in all inertial systems
\bea\label{units}  
d{\hat t}&=&dt \epsilon^\alpha\nonumber\\
d{\hat x}^i&=&dx^i \epsilon^\beta
\eea  
where $\epsilon$ can be any function, and the metric 
(in this case the Minkowski metric) is left unchanged.
If $\alpha=\beta$ we have an active conformal transformation
(for a passive conformal transformation $dx$ and $dt$ are left
unchanged, and the metric is multiplied by $\epsilon^{2\alpha}$).
If $\alpha\neq\beta$, a Lorentz invariant theory is replaced 
by a theory in which $c$ remains isotropic, color independent, and   
independent of the speeds of observer and emitter; but it varies  
like ${\hat c}\propto \epsilon^{\beta-\alpha}$. 
A general unit transformation may be decomposed into
a conformal transformation plus a VSL transformation with $\beta=0$. 

It is immediately obvious that local Lorentz transformations   
in the new units are preserved:  
\bea  
d{\hat t}'&=&\gamma{\left(d{\hat t} - {{\hat v}\over {\hat c}^2}
d{\hat x}
\right)}\nonumber\\  
d{\hat x}'&=&\gamma{\left(d{\hat x}-
{\hat v}d{\hat t}\right)} 
\eea  
with 
\be
\gamma={1\over {\sqrt {1-{\left({\hat v}\over {\hat c}\right)}^2}}}
\ee
Hence the standard definition remains unmodified, if one
employs the local value of $c$  
in the transformation.   
  
A novelty arises because changes of space-time units
do not generally produce new coodinate patches because 
(\ref{units}) needs not be  holonomic: one may have eg.
$d^2{\hat t}\neq 0$.  Hence there would not be 
a global ${\hat t}$ time coordinate: the new  
``coordinate elements'' would not be differentials of any coordinates. 
Even if  in one frame the transformation (\ref{units}) were holonomic,  
in a boosted frame it would not be. Some oddities pertaining 
to the new units follow. Partial derivatives
generally do not commute. The change in the  
``coordinate time'' between two points may depend upon the path taken to link  
the two points.   
  
We have thus identified the structure of a VSL Lorentz invariant theory.  
The theory is locally Lorentz invariant in the usual way,   
using in local transformations the value of $c$ at that point.  
However local measurements of time and space 
are not closed forms, and therefore cannot be made into coordinates. 
Integrating factors can always be found, so that $d{\hat t}/\epsilon^\alpha$
and $d{\hat x}/\epsilon^\beta$ are closed forms, and ${\hat c}=\epsilon
^{\beta-\alpha}$.

Although a time coordinate does not
generally exist,  in many important  
cases it may be defined. 
If $\partial_\mu c\partial^\mu c<0$ then local coordinates
exist so that $c$ only changes in time. We shall call this
the homogeneous frame. Then $d^2 {\hat t}=0$, and a ${\hat t}$ coordinate
can be defined.  Hence if we  
insist upon using a time coordinate we necessarily pick up a preferred 
reference frame - thereby violating the principle of relativity. 
This situation will be true in cosmology (where the preferred
frame is the cosmological frame) but not in the context of static 
solutions, such as black hole solutions. 
Also a time coordinate may always be defined along a line. In particular  
for a geodesic, the amount of proper time is always well defined, although  
the proper time between two points depends on the trajectory (a situation  
already true in general relativity).  

\section{Generalized  covariance} \label{cov1} 
In order to construct a theory of gravitation  
we need to discuss general covariance. 
Covariance is the requirement of invariance under the choice of coordinate  
chart. This may be trivially adapted to VSL if we only use charts   
employing an ``$x^0$'' coordinate, with dimensions of length  
rather than time. Then $c$ appears nowhere in the usual
definitions of differential geometry, which may therefore   
still be used. The laws for the transformation
of tensors are the same as usual. The 
metric is dimensionless in all components   
and does not explicitly depend on $c$; it transforms like a rank 2 tensor.   
The usual Cristoffell connection may be defined from the metric   
by means of the standard formula, without any extra terms in the gradients  
of $c$ (which only appear if we try to revert to a time type of coordinate).  
A curvature tensor may still be defined in the usual way, and a Ricci  
tensor and scalar derived from it. The volume measure does not contain  
$c$.   As we will see, many novelties introduced  
by a varying $c$ only emerge when we try to connect the
$x^0$ coordinate  
with time. 

Whenever applying a unit transformation (\ref{units})
to a covariant theory, the above remarks apply only to the  VSL 
part of the transformation (that is the component with $\beta=0$).  
For the conformal part of the transformation, 
with $\epsilon^{\alpha}=\epsilon^{\beta}=\Omega$,
the structures of differential geometry transform in the usual
way \cite{he}. For instance the Ricci scalar transforms as:
\be
{\hat R}={R\over \Omega^2} -6{\Box \Omega\over \Omega^3}
\ee
for active conformal transformations. 
  
It is not altogether surprising that covariance may be redefined so  
easily for a theory with such different foundations. It has been pointed  
out that covariance is an empty requirement (see \cite{wein,frid}).   
Not only does covariance not imply local Lorentz invariance,  
but also {\it any} theory can be made covariant. An example of a   
covariant formulation of Newtonian gravity is given in \cite{frid}. In this   
theory the tangent space is not a portion of Minkowski space, rather a   
portion of Galilean space.  
 
\section{An overview of the underlying structure}\label{cov2} 
What structure represents covariance and local Lorentz invariance 
when $c$ is allowed to vary? We found that it is a unit-invariant 
redefinition of these concepts, which indeed does not differ much from the  
usual definitions if we phrase them suitably. 
Local Lorentz transformations are 
the same as usual, using the local value of $c$. Covariance and the usual 
constructions of differential geometry remain unchanged  
as long as a $x^0$ coordinate is used, or more generally  
if all coordinates used have the same dimensions.  
 
What is 
new, then? The novelty is that locally made time and space
measurements produce 
a set of infinitesimals which are generally not closed 
forms. Therefore space-time measurements cannot be made into local coordinate 
patches. This leads to the following modification of the structure
of relativity. The underlying structure of general relativity is a manifold,
combined with its tangent bundle (where physics actually happens).
If $c$ varies  the underlying 
structure is a fibre bundle. The base manifold has the same structure 
as usual, but the fibres in which local measurements happen 
are not the tangent bundle. The fibers are vector spaces
obtained by means of a  
non-holonomic transformation over the tangent bundle.  
 
It may seem rather contorted to adopt the above structure when we 
know that a unit transformation would transform it into standard 
covariance and local Lorentz invariance. However such a structure 
has the merit that it only incorporates those elements of the original 
structure which are unit-independent, and can therefore be the  
outcome of experiment. Moreover such structure allows for a varying 
speed of light within a covariant framework, which  
is precluded by the standard framework. What one 
may gain from such extra freedom is a simplified description 
of any given physical situation, when all fine structure 
coupling constants are allowed to vary, in what looks like 
a contrived fashion, if we use units such that $c$ is constant. 
 
We wish to propose a theory which permits space-time 
variations in all coupling constants; more specifically in generalized 
fine structure constants $\alpha_i=g_i^2 /(\hbar c)$ - where $g_i$ 
are the various charges corresponding to all interactions apart from 
gravitation. This purpose draws inspiration from the findings of
\cite{webb}. However we restrict such variations so that 
the ratios between the $\alpha_i$ remain  constant. This suggests 
that attributing the variations in the $\alpha_i$ to changes 
in $c$ or $\hbar$ might lead to a simpler picture.  
In suitable units we could regard our theory as a  
``generalized '' Bekenstein changing $e$ theory, but   
in this system of units the picture is rather contrived.  
 
We will see, in Section \ref{duals}, that a natural dynamics will emerge in  
this theory which becomes unnecessarily complicated when the theory 
is reformulated in fixed $c$ units. Whatever the system of units 
chosen the general theory 
we will consider is not a dilaton theory. Some important geometrical 
aspects (such as inaccessible regions of space-time to be studied  
in Section~\ref{empty}) are missed altogether in the fixed $c$ system.

\section{Matter fields subjected to  VSL}  
\label{matter}  
Before embarking on an investigation of the dynamics of $c$ and  
of gravitation, we first undertake a careful examination of the effects 
of a varying $c$ upon the  
matter fields. The key point here is that it is always possible to  
write the matter Lagrangian so that is does not depend explicitly  
on $c$. We may break this rule, if we wish to,  
but this is not necessary.   
This remark is highly non-trivial, and relies heavily on using   
an $x^0$ coordinate (as opposed to time). The introduction  
of a time coordinate would not only introduce non-covariant elements  
in expressions like $\partial_\mu\phi=(\partial_t\phi/c,\partial_i  
\phi)$, but would also force the matter Lagrangian to  
depend explicitly on $c$, via  kinetic terms.   
  
Using an $x^0$ coordinate the situation is rather different.  
For instance, for a massless scalar field with no interactions we have:  
\be    
{\cal L}_m=-{1\over 2}(\partial_\mu\phi)(\partial^\mu\phi)    
\ee    
which does not depend on $c$. Similarly for a spin 1/2 free massless field   
we have  
\be   
{\cal L}_m=i{\overline \chi}\gamma^\mu\nabla_\mu\chi  
\ee  
The above expressions, in particular the latter, are sometimes  
multiplied by $\hbar c$ (see for example  
Mandl and Shaw \cite{mandl}). If $c$ and $\hbar$ are constant  
this operation has no effects, other than modifying the   
dimensions of the fields. However in a minimal VSL theory  
such an operation should be banned. All dynamical fields should  
be defined with dimensions such that the kinetic terms have no  
explicit dependence on either $c$ or $\hbar$. This forces  
all matter fields to have dimensions of ${\sqrt{E/L}}$.   
  
The only chance for  ${\cal L}_m$ to depend upon $c$ therefore  comes  
from mass and interaction terms. These may always be defined  
so that no explicit dependence on $c$ is present.  
By dimensional analysis this requirement fully defines how masses,   
charges, and coupling constants scale with $c$, provided we know  
how $\hbar$ scales with $c$. This issue will be discussed further 
in the  next Section.

Let us first consider mass terms. For a scalar field we have:  
\be\label{scalarm} 
{\cal L}_m=-{1\over 2}{\left(\partial_\mu\phi\partial^\mu\phi    
+{1\over \lambda_\phi^2}\phi^2\right)}    
\ee    
Hence in a minimal VSL theory  the Compton wavelength   
$\lambda_\phi$ of the particle should   
not depend on $c$. For a massive spin 1/2 particle we have   
\be   
{\cal L}_m=i{\overline \chi}\gamma^\mu\nabla_\mu\chi - {1\over  
\lambda_\chi}{\overline \chi}\chi  
\ee  
with a similar requirement. More generally we find that $c$  
does not appear in mass terms if all particles' masses are proportional  
to $\hbar/c$ (or their rest energies proportional to $\hbar c$).   
  
If we now consider fields coupled to electromagnetism we find that  
the electric charge $e$ should scale like $\hbar c$, if  
explicit dependence on $c$ is to be avoided. Consider  
for instance a $U(1)$ gauged complex scalar field. Its action may  
be written as   
\be  
{\cal L}_m=-(D^\mu\phi)^\star D_\mu\phi  
-{|\phi|^2\over \lambda_\phi^2} -{1\over 4}F_{\mu\nu}F^{\mu\nu}  
\ee  
where the $U(1)$ covariant derivative is  
\be  
D_\mu=\partial_\mu +i{e\over \hbar c}A_\mu   
\ee  
and the electromagnetic tensor is   
\be  
F_{\mu\nu}=\partial_\mu A_\nu - \partial_\nu A_\mu  
\ee  
Hence $e$ should be proportional to $\hbar c$. The same holds true  
for fields of any spin coupled to electromagnetism,   
since $e$ only appears in the definition of the covariant derivative.  
Notice that the constancy of $e/(\hbar c)$ is also required 
for the gauge-invariant field strength tensor not to receive any 
corrections. Gauge transformations should take the form:  
\be  
\delta A_\mu =-{\hbar c\over e}\partial_\mu f
\ee   
for $\delta \phi=i f \phi$, where $f$ is any function. This is  
necessary so that $D_\mu \phi$ transforms covariantly: 
$\delta(D_\mu \phi)=i f D_\mu \phi$. But then the 
gauge invariant field strength tensor must be defined  as  
\be   
F_{\mu\nu}={\hbar c\over e}{\left( \partial_\mu {\left( {e\over \hbar c}  
A_\nu\right)} - \partial_\nu {\left( {e\over \hbar c}  
A_\mu\right)}\right)}  
\ee  
and indeed this receives extra terms if $e/(\hbar c)$ is not constant.  
 
Inspection of the electroweak and strong interaction Lagrangians  
reveals that their coupling charges $g$ should also scale like $\hbar c$.  
This is indeed a general feature for any interaction, and follows  
from dimensional analysis. It is always the combination   
$g/(\hbar c)$ that appears in covariant derivatives and, in   
non-Abelian theories, in the gauge field strength tensor.   
  
Next we discuss two important cases to be used  later  
in this paper: a field undergoing spontaneous symmetry breaking,  
and a matter cosmological constant. Consider a U(1) gauge symmetric 
complex scalar field as above, but with a potential 
\be 
V(\phi)={1\over \lambda_\phi^2}|\phi|^2-{1\over 2\lambda_\phi^2 
\phi_0^2}|\phi|^4 
\ee 
Then the Compton wavelength $\lambda_\phi$ and $\phi_0$ should 
both be independent of $c$. If the quartic term is ignored then the vacuum 
is at $\phi=0$, so that we have a massive complex scalar field (with  
Compton wavelength $\lambda_\phi$), and a massless gauge boson. 
The charge is $e\propto \hbar c$. If we consider the quartic term, 
as is well known, we have spontaneous symmetry breaking. The vacuum 
is now at $|\phi|=\phi_0$. Expanding around the vacuum we find  
a real scalar field with Compton wavelength $\lambda_\phi$, 
and a massive gauge boson with Compton wavelength  
\be 
{1\over \lambda_A}={e\over \hbar c}\phi_0 
\ee 
which therefore is independent of $c$. Hence the rest
energies of all massive particles, 
regardless of the origin of their mass, scale 
like $\hbar c$. Due to spontaneous symmetry breaking  the vacuum 
energy decreases by  
\be 
\Delta V=-{\phi_0^2\over 2\lambda^2} 
\ee 
and so
this process gives rise to a negative vacuum energy, if the 
original vacuum energy is zero. We shall label it by $\Lambda_m 
=\Delta V$, and call it the matter cosmological constant. It
adds a term  to the matter Lagrangian 
\be 
{\cal L}_m=-\Lambda_m 
\ee 
Under minimal coupling $\Lambda_m$ does not depend on $c$.
However we could also allow $\phi_0$, and therefore $\Lambda_m$,
to depend on $c$ (as we shall do in \cite{vcb}). 
 
Finally we consider an example of a classical Lagrangian, that of
a charged particle in a field:  
\bea\label{lagpp} 
&&{\cal L}_m(x^\gamma)=\nonumber\\ 
&&\int d\lambda {\left[- {E_0\over 2} {dy^\mu\over d\lambda}  
{dy_\mu\over d\lambda} + eA_\mu {dy^\mu\over d\lambda}  
\right]} {\delta^{(4)}(x^\gamma -y^\gamma)  
\over {\sqrt {-g}}}  
\eea  
Here the affine parameter is $d\lambda=cd\tau$, where $\tau$ is proper  
time. 
Note that any of the variations sometimes employed in the literature,
eg. using the square root of $-u^2$ (with $u=d x/d\lambda$), should not
be used. This is because, as we shall see, $u^2$ needs not remain  
constant.
Minimal coupling therefore
requires that the particle's rest energy ($E_0=m_0c^2$)   
and charge $e$ be independent of $c$.

In non-minimal theories we may consider a direct dependence on $c$   
in the matter Lagrangian. This is far from new: for instance   
Bekenstein's theory \cite{bek2} allows for a direct coupling   
between a varying $e$ and all forms of matter coupled to   
electromagnetism.

\subsection{A worked out example} \label{mattera}
Consider a massive scalar field $\phi$ in flat space-time 
(metric $\eta_{\mu\nu}={\rm diag}(-1,1,1,1)$ if we use a $x^0$ coordinate)
with a variation in $c$ such that 
\be 
c={c_0\over 1+{c_0t\over R}} 
\ee  
in suitable coordinates (so that $c$ does not vary in space). 
We have defined $c_0$ as the speed of light at time $t=0$. 
At time $t=-R/c_0$ the speed of light goes to infinity. 
As time progresses the speed of light decays to zero,  
as $t\rightarrow\infty$. We shall see that this is indeed the  
solution corresponding to flat space-time. 
Then $\phi$ satisfies: 
\be 
\ddot \phi-\nabla^2\phi+{1\over \lambda_\phi^2}\phi=0 
\ee 
which may be solved with Fourier series, with amplitudes subject to 
\be 
\ddot\phi_k +{\left(k^2+{1\over \lambda_\phi^2}\right)}\phi=0 
\ee 
The solution is  
\be\label{phisol} 
\phi_k=\phi_0(k)e^{i(\pm k^0x^0+{\bf k}\cdot {\bf x})} 
\ee 
with a dispersion relation 
\be  
(k^0)^2=k^2+{1\over \lambda_\phi^2} 
\ee 
As expected there is nothing new if we use a $x^0$ coordinate. 

If we insist on using a time coordinate we find that we can only 
do so in one inertial frame, the one in which the speed of light 
is homogeneous. By requesting to use a time coordinate, and make 
contact with physics, we therefore select a preferred reference  
frame. In this frame: 
\be 
x^0=\int c dt=R\log{\left(1+{c_0t\over R}\right)} 
\ee 
and so we find that around a given time $t=t_0$ we have the  
Taylor expansion: 
\be 
k^0x^0=k^0c(t_0)(t-t_0) +k^0R\log{\left(1+{c_0t_0\over R}\right)} 
\ee 
We find that the local  
frequency changes proportionally to $c$: 
\be 
\omega(t)=k^0c(t)={k^0c_0\over 1+{c_0t\over R}} 
\ee 
In addition there is a phase shift with value 
\be 
\Phi_0=k^0R\log{\left(1+{c_0t_0\over R}\right)} 
\ee 
As we approach the initial singularity the wave 
suffers infinite blueshift.
As time flows it redshifts progressively. The 
similarities between this effect and the cosmological redshift 
have been pointed out in \cite{man}. However the effect presented here 
is not due to gravity (expansion) but is due purely to the varying  
speed of light.  
 
Naturally the above identification of a local frequency 
is only valid if $\omega \gg |\dot c/c|$. This amounts to 
requiring: 
\be 
k^0\gg {1\over R{\left(1+{c_0t\over R}\right)}} 
\ee 
Hence any  plane-wave approximation breaks down
near the initial singularity; an interesting result.


\section{Quantization } \label{quanta} 
Unfortunately the requirement that $c$ does not appear explicitly  
in ${\cal L}_m$ does not fix the scaling with $c$ of all ``constants'':  
Planck's and Boltzmann's constants, $\hbar$ and $k_B$, are left
unfixed. Furthermore these two ``constants'' cannot be fixed by the  
classical dynamics, i.e. by adding to the action dynamical terms 
in two scalar fields 
$\hbar$ and $k_B$ (as we shall do with $c$).  Instead these  
have to be provided as a function of $c$,  by means of  
scaling laws $\hbar(c)$ and $k_B(c)$. These scaling laws 
should be regarded as postulates of the theory. 
 
Here we explore the implications of various $\hbar(c)$ laws. 
Let us consider first the simple case of 
a non-relativistic linear harmonic  
oscillator. Its Lagrangian is given by:  
\be  
L=\frac{1}{2}m\dot x^2 -\frac{1}{2}m\omega^2 x^2  
\ee  
and we assume that $m$ and $\omega$ are independent of $c$,  
and therefore constant. Its Hamiltonian is 
\be   
H={p^2\over 2m}+{m\omega^2 x^2\over 2} 
\ee  
and is time-independent. We postulate that quantization produces
an expression of the form
\be
{\hat H}=\hbar\omega({\hat N} +1/2  
)  
\ee 
where ${\hat N}$ is the particle number operator, ie: the classical 
energy of the oscillator is in quanta of energy $\hbar \omega$.
Hence
\be\label{neqn}
{d\over dt}\hbar({\hat N} +1/2)=0
\ee
This implies that should $\hbar$ drop, the particle number
would increase, which is hardly surprising.
Indeed the amplitude $A$ and frequency 
$\omega$ of the classical  
oscillations remain constant, and therefore so does their total  
energy $E=m\omega^2 A^2/2$. However the quantum particles  
contained in the oscillator have energies $\hbar \omega$ which vary  
like $\hbar$. To reconcile these two facts the number of particles  
has to vary, proportionally to $1/\hbar$.  
Such a phenomenon has a clear experimental meaning, since the number  
of particles does not depend on the units being used. 

Furthermore if the oscillator is initially in the vacuum state, a drop
in $\hbar$ suppresses the zero-point energy. Particles should
therefore be produced  so that ${\hat H}$
remains constant. 
We have both particle multiplication and particle
production (a phenomenon noticed before in VSL theories by \cite{harko}).

Since creation and anihilation operators satisfy a time-independent
algebra $[a,a^\dagger]=1$, the best way to express the variability
of ${\hat N}$ is by means of a Bogolubov-type  transformation. A short
calculation shows that:
\be
a(t')=\alpha a(t) + \beta^\star a^\dagger(t)
\ee
with 
\bea
|\alpha|^2&=&{\hbar(t)+\hbar(t')\over \hbar(t')}\\
|\beta|^2&=&{\hbar(t)-\hbar(t')\over \hbar(t')}
\eea
enforces that the expectation values of ${\hat N}(t)=a^\dagger
(t)a(t)$ satisfy
(\ref{neqn}). 

This discussion generalizes to relativistic quantum 
field theory, with $\hbar c$ replacing $\hbar$. 
Now we should have: 
\be 
{\hat H}=\sum_{\bf k}\hbar \omega ({\hat N}+1/2) 
\ee 
with $\omega=k^0 c\propto c$. Hence now 
\be
{d\over dx^0}\hbar c ({\hat N} +1/2)=0
\ee
The time dependence in ${\hat N}$
can now be expressed in the form of a Bogolubov transformation 
\be
a(k^0,x^{\overline \mu})=\alpha a(k^0,x^\mu) + \beta^\star 
a^\dagger(k^0,x^\mu)
\ee
with 
\bea
|\alpha|^2&=&{\hbar(x^\mu) c(x^\mu)+\hbar(x^{\overline \mu}) c(x^{\overline \mu})\over 
\hbar(x^\mu) c(x^\mu)}\\
|\beta|^2&=&{\hbar(x^\mu) c(x^\mu)-\hbar(x^{\overline \mu}) c(x^{\overline \mu})\over 
\hbar(x^\mu) c(x^\mu)}
\eea

Recalling that
$g_i/(\hbar c)$ is a constant, we have particle production 
at a rate proportional to  $
1/\alpha_i$, where $i$ labels the various interactions. 
We shall parameterize $\hbar(c)$ 
by means of an exponent $q$ such that  
\be\label{alphan} 
\alpha_i\propto g_i\propto \hbar c\propto c^{q} 
\ee

\section{Gravitational  dynamics} \label{grav} 
We now set up some possibilities for actions governing the 
evolution of the metric and  speed of light. Only a small class
of these actions may be transformed into a dilaton action, by means 
of a unit transformation.  In Appendix we describe a somewhat 
orthogonal approach. 
 
We shall take as our starting point the action of General Relativity:  
\be \label{SGR} 
S= \int d^4x \sqrt{-g}{\left(R-2\Lambda  
+{16\pi G\over c_0^4 }{\cal L}_m \right)}  
\ee  
where $R$ is the Ricci scalar, and $\Lambda$ is the geometrical  
cosmological constant (as defined in \cite{maeda,am}) and ${\cal L}_m$  
is the Lagrangian of all the matter fields (including the
above mentioned matter cosmological constant).  
 
A changing $G$ theory was proposed by Brans and Dicke \cite{dick2}, and 
we shall work in analogy  
to this generalization of General Relativity  
in what follows, albeit with a couple of 
crucial differences.  The idea in this paper (in \cite{dick2}) is to  
replace $c$ ($G$) by a field, wherever it appears in (\ref{SGR}). 
In addition one should add a term to the Lagrangian describing 
the dynamics of $c$ ($G$). An ambiguity appears because (\ref{SGR})  
may be divided by any power of $c$ ($G$), before the replacement is  
performed. Brans and Dicke avoided commenting on this ambiguity,  
and cunningly performed the necessary division by $G$ which led  
to a theory with energy conservation. We shall not be hampered by this 
restriction; indeed we expect violations of energy conservation 
in VSL. Hence we consider actions in which the replacement 
is made after the most general division by $c$ is made. 
In the simplest case we define a scalar field  
\be  
\psi=\log{\left( c\over c_0 \right)}  
\ee  
so that $c=c_0e^\psi$, and take 
\be\label{S21}    
S= \int d^4x \sqrt{-g}( e^{a\psi}( 
R-2\Lambda +{\cal L}_{\psi}) 
+{ 16\pi G\over c_0^4}e^{b\psi}{\cal L}_m ) 
\ee 
The simplest dynamics for $\psi$ derives from: 
\be 
{\cal L}_{\psi}=-\kappa(\psi) \nabla_\mu\psi\nabla^\mu\psi  
\ee 
where $\kappa(\psi)$ is a dimensionless coupling function (to be 
taken as a constant in most of what follows). 
We shall impose $a-b=4$, although this is not necessary.

Notice that $a=4$, $b=0$, is nothing but a unit transformation 
applied to Brans-Dicke theory, with 
\bea  
\phi_{bd}&=&{e^{4\psi}\over G}\\  
\kappa(\psi)&=&16\omega_{bd}(\phi_{bd})  
\eea 
This shall be proved in Section~\ref{duals}, where we identify the full set of 
cases which are a mere unit transformation applied to existing 
theories. Among the theories which are truly new,  
$a=0$, $b=-4$ is particularly simple and we shall call it 
minimal VSL.  
  
We can trivially generalize this construction, 
by complicating the dynamics encoded in ${\cal L}_{\psi}$, for instance 
by adding a potential $V(\psi)$ to it. We can 
also take for $\psi$ a complex, vector, or spinor field, with 
the speed of light deriving from a scalar associated with $\psi$ 
(eg. ${\overline\psi} \psi$ for a spinor field).  A nice example  
(developed further in Section~\ref{ftracks}) is a theory in which $\psi$ is  
a complex field, with  
\be 
c=c_0e^{-|\psi|^2} 
\ee 
and with a Mexican hat potential added to ${\cal L}_{\psi}$. 
 
Another important novelty of our theory, not included in Brans-Dicke 
theory (but noted by \cite{maeda}), is that we allow $\Lambda$ and  
$\Lambda_m$ to depend on $c$. It seems fair to allow $\Lambda$,  
like $\hbar$ or $k_B$, to depend on $c$. After all $\Lambda$ is  
a much less fundamental constant.  On the contrary 
if $\Lambda_m$ depends on $c$, then so does the vacuum expectation
value $\phi_0$, and so we have gone beyond minimal matter coupling. 
In what follows we shall absorb $\Lambda_m$ into a total geometrical  
Lambda 
\be  
{\overline \Lambda}=\Lambda+{8\pi G\over c^4}\Lambda_m 
\ee 
In our applications to cosmology \cite{vcb} we shall assume that
\be \ 
\Lambda\propto (c/c_0)^n=e^{n\psi} 
\ee 
and  
\be 
\Lambda_m\propto (c/c_0)^{m}=e^{m\psi} 
\ee 
We will see that allowing $\Lambda$ to depend on $c$ leads 
to interesting 
cosmological scenarios \cite{vcb}. In such theories it is the presence 
of a Lambda problem that drives changes in the speed of light. 
These in turn solve the cosmological constant and other problems 
of Big Bang cosmology. In effect Lambda acts as 
a potential driving $\psi$.


\subsection{Gravitational field equations} 
The field equations in this theory may now be derived by varying the 
action. Variation with respect to the metric leads to  
gravitational equations  
\bea \label{einstein21}    
G_{\mu\nu}+\Lambda g_{\mu\nu}&=&{8\pi G\over c^4}    
T_{\mu\nu}+ \kappa{\left(\nabla_\mu\psi \nabla_\nu\psi    
-{1\over 2}g_{\mu\nu}\nabla_\delta\psi \nabla^\delta\psi  
\right)}   \nonumber \\ 
&&+e^{-a\psi}(\nabla_\mu\nabla_\nu e^{a\psi} 
-g_{\mu\nu}\Box e^{a \psi})   
\eea  
where the matter stress energy tensor is defined as usual:    
\be    
T_{\mu\nu}={-2\over \sqrt{-g}}{\delta S_m\over \delta g^{\mu\nu}}    
\ee    
These equations  are particularly simple for minimal VSL
($a=0$ and $b=-4$).
 
Variation with respect to $\psi$ leads to  
\bea\label{psi21}  
\Box \psi&&  +a\nabla_\mu\psi \nabla^\mu\psi\nonumber\\ 
&&={8\pi G\over c^4(2\kappa + 3a^2)} 
(a T -2a\rho_\Lambda -2b 
{\cal L}_m ) +{1\over \kappa}{d{\overline 
\Lambda}\over d\psi} 
\eea  
Again minimal VSL is particularly simple:
\be\label{psi2.1}  
\Box \psi  
={32\pi G\over c^4\kappa}{\left({\cal L}_m -
{\left(1-{m\over 4}\right)}\Lambda_m\right)} +{1\over \kappa} 
n\Lambda 
\ee 
As announced above,  
in general either a matter or a geometrical Lambda 
drive changes in $c$. The total matter Lagrangian
${\cal L}_m$ also drives changes in $c$, if $b\neq 0$. Ambiguities
in writing ${\cal L}_m$ (total divergences) are therefore
relevant for $c$, as indeed for the matter field equations
under VSL (see below).

\subsection{Impact upon matter field equations}
Bianchi identities applied to (\ref{einstein21}) and
(\ref{psi21}) imply 
\be \label{bianchi}
\nabla_\mu (T^\mu_\nu e^{b\psi})=b e^{b\psi}
{\cal L}_m\nabla_\nu\psi  
\ee  
or equivalently:
\be
\nabla_\mu T^\mu_\nu =-b (T^\nu_\mu -\delta^\nu_\mu{\cal L}_m)
\nabla_\nu\psi 
\ee 
Therefore we only have energy conservation if $a=4,$ 
$b=0$. In all other cases a varying $c$ creates or 
destroys energy; indeed beyond the naive expectation 
(the term in ${\cal L}_m$ is far from expected). 
This fact merely reflects the interaction
between the matter fields and the gravitational field $\psi$, 
present due to the coupling $e^{b\psi}{\cal L}_m$. This
interaction affects the field equations for matter, beyond
what was descrobed in Section~\ref{matter} (which is only strictly correct 
if $b=0$).
Indeed taking the variation with respect to matter
fields, in every situation where it is usual to 
 neglect a full divergence, a new
term in $\partial^\mu\psi$ now appears. 
For instance scalar fields satisfy a modified Klein-Gordon equation:  
\be     
{\left(\Box -{1\over \lambda_\phi^2}\right)}\phi=
-b\nabla_\mu\phi\nabla^\mu  
\psi    
\ee    
with gradients of $\psi$ driving the field $\phi$ 
and therefore changing its energy balance. All field
equations will be similarly affected, with a net result
that energy conservation is violated according to 
(\ref{bianchi}). 

To give a concrete example, the plane wave solution studied in 
Section~\ref{mattera}
is now subject to:
\be 
\ddot\phi_k +{\left(k^2+{1\over \lambda_\phi^2}\right)}\phi=
-b{\dot\phi_k\over R}
\ee 
A solution is
\be\label{phisol1} 
\phi_k=[\phi_0(k)e^{-{bx^0\over R}}]e^{i(\pm k^0x^0+{\bf k}\cdot {\bf x})} 
\ee 
subject to the same dispersion relation. Hence, in addition to the effects
studied in Section~\ref{mattera}, 
the amplitude of the plane waves is now proportional
to $c^b$. If $R>0$, and $b>0$, we not only have a ``redshift effect''
(affecting the energy of the field quanta), but the classical energy 
of the field also dissipates.

Finally note that
we may also take on board terms which are usually neglected in minimal 
theories because they are full divergences. If $b\neq 0$
these terms affect the matter field equations;
indeed they drive changes in $c$. For instance
one could consider electromagnetism based on
\be
{\cal L}_m=-{1\over 4}(F_{\mu\nu}F^{\mu\nu}
+\zeta F_{\mu\nu}{\tilde F}^{\mu\nu})
\ee
where ${\tilde F}$ is the dual of $F$ and $\zeta$ is a constant.
The second term is usually irrelevant, because it is a full
divergence. However we now have Maxwell's equations:
\be
\nabla_\mu F^{\mu\nu}+ b (F^{\mu\nu}+\zeta {\tilde F}^{\mu\nu})
\partial_\mu\psi=j^\nu
\ee
where $j^\nu$ is the electric current. 

\subsection{Effect upon classical particles} 
These processes are also reflected in the equations of motion 
for a point particle. From (\ref{lagpp}), with $e=0$,
we can derive the 
stress energy tensor: 
\be  
T^{\mu\nu}(x^\delta)=mc^2\int d\lambda {dy^\mu\over d\lambda}  
{dy^\nu\over d\lambda} {\delta^{(4)}(x^\delta -y^\delta(\lambda))  
\over {\sqrt {-g}}}  
\ee  
where we have assumed that $mc^2$ is a constant (so that the matter 
Lagrangian does not depend on $c$). From (\ref{bianchi})
one gets:
\be \label{neogeo} 
{d^2 x^\mu\over d\lambda^2}+\Gamma^\mu_{\nu\delta}   
{dx^\nu\over d\lambda}{dx^\delta\over d\lambda}
=-b{\left({dx^\mu\over d\lambda} {dx^\nu\over d\lambda}
-{1\over 2}{dx^\alpha\over d\lambda}{dx_\alpha\over d\lambda}
 g^{\mu\nu}\right)}\psi_{,\nu} 
\ee  
where we recall $d\lambda=cd\tau$. Alternatively we may integrate
the volume integral in (\ref{lagpp}) to obtain action: 
\be \label{neogeo1} 
S=-{E_0\over 2}\int d\lambda e^{b\psi}g_{\mu\nu}\dot x^\mu\dot x^\nu 
\ee 
Direct variation of this action is equivalent to 
equation (\ref{neogeo}) and may be more practical. 
An immediate first integral of this action is:
\be 
u^2=u^2_0(c/c_0)^{-b} 
\ee 
with $u^\mu=dx^\mu/d\lambda$.
Hence null particles remain null, but time-like lines have a variable 
$u^2$.

We see that matter no longer follows geodesics. However all bodies with the 
same set of initial conditions fall in the same way. A weak 
form of the equivalence principle is therefore satisfied.  
In particular there is no conflict between these theories and
the Eotvos experiment. 
In \cite{vcc} we shall investigate the impact of these effects upon
the standard tests of gravitational light deflection, and the
perihelium of Mercury.  Here, however, we limit ourselves to
integrating the geodesic equation in the local free-falling frame,
or in flat space-time. Then (\ref{neogeo1}) produces the 
Lagrangian
\be
{\cal L}=e^{b\psi}(-\dot{(x^0)}^2+\dot x^2)
\ee
where dots represent $d/d\lambda$. There are three conserved 
quantities: $E=\dot x^0e^{b\psi}$, $p=\dot x e^{b\psi}$,
and ${\cal L}=-1$, from which we may conclude
\be\label{geoint}
{v^2\over c^2}={p^2\over E^2}=1-{e^{b\psi}\over E^2}
\ee
As a result the  particle's gamma factor
\be
\gamma^2={1\over 1-v^2/c^2}\propto c^{-b}
\ee
If $b\neq 0$ the field $\psi$ will therefore accelerate
or brake particles.

\section{Fixed speed of light duals}\label{duals} 
We now identify which of our theories  
are simply well-known fixed $c$ theories subject to a change of units.  
By doing so we will also expose the undesirable complication
of the fixed $c$ picture  in all other cases.

Let us first rewrite our theories in units in which $c$, $\hbar$, 
and $G$ are fixed, but the couplings $g$ are variable, thereby
mapping VSL theories into ``Bekenstein'' changing charge 
theories. Recalling that in VSL units we have $\alpha_i\propto g_i\propto 
\hbar c\propto c^{q}$ (cf. Eqn.~(\ref{alphan}), we should  
perform the following change of units: 
\bea 
{d\hat t}&=&dt e^{(3-{q \over 2})\psi}\\ 
{d\hat x}&=&dx e^{(2-{q \over 2})\psi}\\ 
{d\hat E}&=&dEe^{(-2-{q \over 2})\psi} 
\eea 
In the new units ${\hat c}$, ${\hat \hbar}$, and ${\hat G}$ 
are constant, ${\hat g}\propto e^{{q\over 2}\psi}$, 
and indeed ${\hat \alpha}=\alpha\propto c^{q}$. Subjecting  
a VSL minimally coupled matter action to this transformation  
leads to an action very far from minimal coupling. Indeed 
all matter fields, eg $\phi$, transform like 
\be 
{\hat \phi}=\phi e^{-2\psi} 
\ee 
Hence all kinetic terms become rather 
contorted, since in the new units
\be 
\partial_\mu\phi\rightarrow\partial_{\hat\mu}{\hat \phi} 
+2{\hat \phi}\partial_{\hat\mu}\psi 
\ee 
This leads to complex additions
to mass and interaction terms. 
For gauged fields we have  
\be 
D_\mu\phi\rightarrow D_{\hat\mu}{\hat\phi}+2{\hat \phi} 
\partial_{\hat\mu}\psi 
\ee 
leading to similar complications, and to 
breaking of standard gauge invariance. 
In conclusion we can transform a VSL theory which is minimally  
coupled to matter (up to the $b\neq 0$ factor) into a fixed 
$c$, $\hbar$ and $G$ theory. However the result is a rather  
unnatural construction, quite distinct from the changing
charge theories previously discussed.  None of our theories is  
a standard changing charge theory in disguise; indeed choosing
a standard changing $g_i$ picture for them is undesirable.

The above may be avoided if we map our VSL theories into 
theories in  which $c$ and $\hbar $ are constants, but $G$ may vary. 
Then in order to preserve minimal coupling all matter fields should 
remain unaffected by the unit transformation, eg ${\hat \phi}=\phi$
This requires  
\bea 
{d\hat t}&=&dt e^{(1-{q \over 2})\psi}\\ 
{d\hat x}&=&dx e^{-{q \over 2}\psi}\\ 
{d\hat E}&=&dEe^{-{q \over 2}\psi} 
\eea 
and so we have that  
\be 
{{\hat G}\over G}=e^{-4\psi} 
\ee 
While this ensures minimal coupling for all quantum fields,
it does not do the job  for classical point particles, if
$q\neq 0$. With the above change of units
one should make the identification
\be 
{\hat \phi}_{bd}={1\over {\hat G}}=e^{4\psi} 
\ee 
and write down the transformed action: 
\bea\label{S21trans}    
{\hat S}&=& \int d^4{\hat x} \sqrt{-{\hat g}}({\hat \phi}_{bd}^{a+q
\over 4}(
{\hat R}-2{\hat \Lambda}) \nonumber\\
&& -{f({\hat \phi}_{bd})\over
{\hat \phi}_{bd}}\nabla_\mu{\hat \phi}_{bd}\nabla^\mu{\hat \phi}_{bd}
+{ 16\pi G_0\over c_0^4}{\hat \phi}_{bd}^{b+q\over 4}{\hat{\cal L}}_m ) 
\eea 
We see that only theories for which $b+q=0$ are scalar-tensor theories 
in disguise. If $q=0$ all structure ``constants'' $\alpha_i$ 
are constant, and indeed 
for $b=0$ (and so $a=4$) we can recognize in ${\hat S}$
the Brans-Dicke action. 
However we see that there are also changing $\alpha$ theories which are 
really  Brans Dicke theories in unusual units: theories with 
$b=-q\neq 0$. Such theories are dilaton 
theories. On the contrary, if $b+q\neq 0$ we have theories 
which can never be mapped into dilaton theories.  

In addition one may perform conformal transformations upon VSL
theories, mapping them into other VSL theories with different 
$a$ and $b$. The relevant formulae shall be given in
\cite{vcc}. By means of conformal transformations it is always
possible to write action (\ref{S21}) as a scalar-tensor theory,
if the matter Lagrangian is homogeneous in the metric. 
The latter, however, is clearly not true, carrying with it the crucial 
implication that there is only
one frame in which the coupling to matter is of the form
$e^{b\psi}{\cal L}_m$, with ${\cal L}_m$ independent of $\psi$.
Thus, the much heralded equivalence between conformal frames is
broken as soon as matter is added to gravity and $\psi$ (a point 
clearly made by \cite{qui}). One may recognize $a=4$, $b=0$ as the Jordan's
frame, $a=0$, $b=-4$ as the Einstein's frame, and $a=b=1$ as 
the tree-level string frame. We should also note that the general 
class of couplings we have considered is contained within the theories 
proposed by Damour and Polyakov \cite{damour} as representing
low-energy limits to string theory, beyond tree-level. More
specifically, using the notation of \cite{damour}, our theories
are those for which $B_i(\Phi)$ is the same for all
the matter fields.

In spite of these comments, in \cite{vcc} we shall make 
use of conformal transformations to isolate the geodesic frame: the frame 
in which free-falling charge-free particles follow geodesics. This can 
always be defined because the Lagrangian of these particles is indeed 
homogeneous in the metric. While this trick simplifies some calculations, 
one should always bear in mind that the geodesic frame only looks simpler
because a lot of gargabe is swept under the carpet by not
writing the Lagrangian for all the other matter fields. 
Minimal coupling to all forms of matter always picks up a 
preferred frame, which is not the geodesic frame unless $b=0$.

\section{Empty space-time}\label{empty}  
The analogue of Minkowski space-time may be derived by setting 
$T^\mu_\nu=\Lambda=0$ in  Equation (\ref{einstein21}). We should also 
set $a=0$ and $\kappa=0$ so as to switch off the gravitational
effects of $\psi$.  Then
$g_{\mu\nu}=\eta_{\mu\nu}={\rm diag }(-1,1,1,1)$, 
using an $x^0$ coordinate.

The speed of light can be found from (\ref{psi21}), which
in coordinates in which $\psi$ is homogeneous becomes
$\ddot \psi=0$. This leads to $\dot\psi={1\over R}$,
where $R$ is an integration constant with dimensions of length
($R$ can be positive or negative). 
If we only use coordinates in which $\psi$ is homogeneous (or, as 
we shall see, if we stay close to the origin compared to the distance $R$) 
then a global time coordinate $t$ may be defined. In terms of it we have: 
\be  
{1\over c^2}{dc\over dt}={1\over R}  
\ee  
which integrates to   
\be  
c={c_0\over 1 +{c_0 t\over R}}  
\ee  
This is nothing but $c$ near the
origin in Fock-Lorentz space-time, in which 
\be
c({\bf r},t;{\bf n})={c_0\over 1+{c_0 t\over R}}  {\left({\bf n}
+{{\bf r}\over R}\right)}
\ee
Even though global coordinates cannot be generally defined
if $c$ varies,  we find that this case is special.
Relations
\bea\label{fock1}
d{\hat t}&=&{dt \over {\left(1 +{c_0 t\over R}\right)}^{N+1}}\nonumber\\
d{\hat r}&=&{d{\bf r} \over {\left(1 +{c_0 t\over R}\right)}^{N}}
\eea 
(corresponding to $\alpha=N+1$ and $\beta=N$)
may be recovered from 
\bea 
{\hat t}&=&{R\over N c_0}{\left(1-{1\over {\left(1+{c_0t\over R}
\right)}^N}
\right)}\\ 
{\hat {\bf r}}&=&{{\bf r}\over {\left(1+{c_0 t\over R}\right)}^N} 
\eea 
(with $N>0$) as long as $|{\bf r}|\ll|R|$. Hence, near the origin there are
global varying $c$ coordinates $t$ and ${\bf r}$. Global transformation
laws between inertial frames may be derived for these coordinates
by writing global Lorentz transformations
for ${\hat t}$ and ${\hat {\bf r}}$ and then re-expressing them
in terms of $t$ and $ {\bf r}$.

The case $N=1$ is particularly simple. It corresponds to $q=2$
for a fixed $G$ representation ($\alpha\propto c^2$), or $q=-2$
in a minimal varying $G$ representation ($\alpha\propto 1/c^2$).
In these cases we have global Lorentz transformations for 
coordinates 
\bea 
{\hat t}&=&{t\over 1+{c_0t\over R}}\\ 
{\hat {\bf r}}&=&{{\bf r}\over 1+{c_0 t\over R}} 
\eea 
These can be inverted into transformations for $t$ and $ {\bf r}$:
\bea 
t'&=&{\gamma\left(t-{{\bf v}\cdot {\bf r}\over c_0^2}\right) 
\over 1-(\gamma -1){c_0t\over R} +\gamma{{\bf v}\cdot {\bf r} 
\over Rc_0}}\\  
{\bf r}_{||}'&=&{\gamma\left({\bf r}_{||}-{{\bf v}\over c_0}t\right) 
\over 1-(\gamma -1){c_0t\over R} +\gamma{{\bf v}\cdot {\bf r} 
\over Rc_0}}\\  
{\bf r}_\perp'&=&{{\bf r}_\perp 
\over 1-(\gamma -1){c_0t\over R} +\gamma{{\bf v}\cdot {\bf r} 
\over Rc_0}} 
\eea 
where $v$ is the velocity between two inertial frames at
the origin at $t=0$ (the velocity 
between two inertial frames varies in space
and time and is proportional to $c$ \cite{man,step}). The transformation
we have just obtained is  the Fock-Lorentz transformation.

This is an interesting result! The Fock-Lorentz transformation 
was first derived by Vladimir Fock in his textbook \cite{fock} as a  
pedagogic curiosity. Special relativity may be derived from 
two postulates: the principle of relativity and the principle of constancy 
of the speed of light. The latter may be replaced by the requirement 
that the transformation be linear. Fock examined the effects of dropping  
the second postulate while keeping the first. He thus arrived at 
a fractional linear transformation identical with the one we have 
just derived.  
 
We have just produced an alternative derivation, based on our 
dynamical equations for the field $\psi$. The constant $R$ in 
the Fock Lorentz transformation appears as an integration constant 
in our solution. Some features of the Fock transformation, not
accommodated by our theory (such as an anisotropic $c$), can be
neglected if we stay close to the origin. Similarly some features
of our theory not present in Fock's theory (such as non-integrability
of infinitesimals) can be ignored in the same region. Hence it is
not surprising that we have arrived at the same construction. 

The Fock-Lorentz transformation has a number of interesting properties,
and one of them is crucial for understanding VSL theories. If we
consider a proper time interval $\Delta t_0$ (referred to the origin)
we find that this is seen in the lab frame as
\be
\Delta t={\Delta t_0 \over (1+c_0\Delta t_0/R)\gamma -c_0\Delta t_0/R}
\ee
which is qualitatively very different from the usual twin
paradox expression.  In the standard theory the only
invariant non-zero time lapse is infinity. In  Fock's theory
such a role is played by $\Delta t_0=-R/c_0$; in contrast infinity
is no longer invariant but can mapped into finite times and vice-versa.
Suitable particle life-times may be mapped to infinity (ie:
stability) by a Fock-Lorentz transformation.

Closer inspection shows that if we look at these theories
from a fixed $c$ perspective $t=-R/c_0$ is indeed mapped into
${\hat t}=\infty$ for $R<0$ (or ${\hat t}=-\infty$ for $R>0$).
This is obvious from (\ref{fock1}) but also true for other values
of $N$. Given that the two representations are globally very 
different one must ask  which representation is more physical.

\subsection{Interaction clocks and trans-eternal times} 
Clearly a change of units transforms our construction into 
plain Minkowski space-time.  Then why not use the fixed $c$ 
representation? The point is that the correspondence is only local.
We can extend the VSL empty solution beyond $t=t_{max}=-R/c_0$, for 
$R<0$. Such extension corresponds  to extending Minkowski space-time 
beyond $t=\infty$. The choice between the two representations
is therefore dependent on whether this extension is physical
or not. 

Let us first examine $t\rightarrow t_{max}$
in units in which $c$  varies. In this picture $c$
goes to infinity at $t_{max}$; but this has implications
on the  time-scales of processes mediated by all interactions.
Decay times, rates of change, etc, all depend on the $\alpha_i$. 
A typical time scale associated with a given interaction 
with energy $Q$  is 
\be \label{tau}
\tau={\hbar\over \alpha^2 Q} 
\ee 
In a minimal VSL theory $Q\propto \hbar c\propto c^q$, $\alpha\propto c^q$,
and so $\tau\propto 1/c^{2q+1}$. 
But our sensation of time flow derives precisely from change,
and this is imparted by interactions and their rates. One may therefore argue 
that a more solid definition of time should be tied to the 
rates $\tau$, and that a more physical clock should 
be obtained by making it tick to $\tau$.  
Like all other definitions of time, this definition should 
not affect physics (which is dimensionless); however it may lead 
to a clearer picture.  

In the varying $c$ picture,  
the number of cycles of an interaction clock  
as $t\rightarrow t_{max}$  is 
\be 
\int^{t_{max}} {dt\over \tau} 
\ee 
which converges if $q<0$, that is if all $\alpha_i$ go to zero
(all interactions switch off). 
Hence our claim that the space is extendable beyond $t=t_{max}$ 
is physically meaningful, if $q<0$.

Let us now examine the same situation in fixed $c$ units.
Even though $c$ and $\hbar$ are now fixed, this is not really
just Minkowski space-time. At the very least all charges $g_i$ 
must now be variable, to produce the same changing $\alpha_i$.
If we want to keep all parameters in (\ref{tau}) constant
except for the $\alpha_i$ we should change units in the following 
way:
\bea 
{d\hat t}&=&dt e^{(1-2q)\psi}\\ 
{d\hat x}&=&dx e^{-2q \psi}\\ 
{d\hat E}&=&dEe^{-q \psi} 
\eea 
For $q<0$ we have that $t_{max}$ is indeed mapped into ${\hat t}=\infty$.
However we find that the number of ticks of an interaction clock 
as we approach ${\hat t}=\infty$
\be 
\int^{\infty}{d{\hat t}\over {\hat \tau}} 
\ee 
converges. Hence the temporal infinity of ``Minkowski'' space-time  
in this theory is spurious. Any natural process would 
slow down as ``fixed-$c$ time'' went on. More and more of this 
``time'' would be required 
for any interaction process to take place. Given that our sensation of time 
flowing is attached to these processes, we could claim that conversely we 
would feel that ``fixed-$c$'' time would start to go faster and faster. 
The fact that a finite number of physical ticks is required to reach 
${\hat t}=\infty$ means that any observer could in fact flow through 
eternity. Such  Minkowski space-time is physically extendable 
beyond $t=\infty$.
 
We have found the first example of a situation 
in which the fixed $c$ representation, while locally equivalent to 
a varying $c$ representation, may be globally misleading.  
The advantage of varying $c$ units in this case is that they
locate at a finite time distance  what can in fact be reached within a finite  
number of cycles of an interaction clock.

  

\section{Fast-tracks in VSL flat-space}\label{ftracks}
More fascinating still is the existence of high-$c$ lines, 
which we shall call fast-tracks. These are flat space-time solutions, 
in theories in which $\psi$ is driven by a potential.  
We first establish the possibility of such  
solutions. Let $\psi$ be a complex scalar field, with a $U(1)$ 
symmetry which may or may not be gauged (we assume it's gauged in what
follows). Let the speed of light 
be given by $c=c_0e^{-|\psi|^2}$. With these modifications we also
have to modify the terms in $a$ and $b$ in (\ref{S21}), but not if $a=b=0$,
as we shall assume. Let us also assume that the field is driven by
a potential
\bea
{\cal L}_\psi&=&-(D_\mu\psi)^\star(D^\mu\psi)-V(\psi)\\
V(\psi)&=&{1\over \lambda_\psi^2}(|\psi|^2-\psi_0^2)^2 
\eea 
where $\psi_0$ is the field's vacuum expectation value,
and $\lambda_\psi$ is the Compton wave-length of $\psi$.

Let us consider a Nielsen-Olesen vortex solution to this 
theory, that is a solution with a boundary condition: 
\be 
\psi=\psi_0e^{in\theta} {\qquad \rm as \qquad} r\rightarrow\infty 
\ee 
Such a solution is topologically stable. In the vortex's core, 
$|\psi|\approx 0$ and so the speed of light is $c_0$. The speed of light 
outside the core (which is $c_0e^{-\psi_0^2}$) is therefore much  
smaller. The field $\psi$ undergoes spontaneous 
symmetry breaking and the unbroken phase, realized in the string's 
core, displays a much larger speed of light.  
An approximate solution for $r\rightarrow
\infty$  is
\be
\psi=(\psi_0+e^{-r/\lambda_\psi})e^{in\theta}
\ee
Hence the string core has a width of order $\lambda_\psi$, 
which could easily be macroscopic; outside the core variations
in the speed of light die off rapidly. The jump in the speed of
light is exponential and depends only on $\psi_0$. For 
$\psi_0\approx 3$, say, the speed of light could be ten orders
of magnitude faster inside the string's core. The size of the
core, and the jump in $c$, are related to independent parameters.

What would happen if an observer travelled along the string, inside
its core? 
Let a cylinder of high-$c$ connect two distant galaxies. Then inside  
the tube $v\propto c$ (cf. (\ref{geoint}) with $b=0$).
Let us assume that $v\ll c$ so that no relativistic effects are present.  
Then the observer could move very fast between these two galaxies, 
returning without any time dilation effects having taken place. 
There would not be a twin paradox - clearly this situation, if realizable, 
is just what intergalactic travel is begging for. 
In practice, to avoid different aging rates between  
sedentary and the nomadic twins we should keep 
the aging pace $\tau$ fixed, ie: $q=-1/2$. Furthermore
in order for the $x^0$ coordinate to track proper-time for
all observers we should have $\alpha=0$ (this point will be developed
further in \cite{vcc} in connection with radar echo delay experiments).

In a dual representation, in fixed $c$ units, fast tracks are
wormholes. If $\tau$ is to remain unchanged, and if $c$ is to be
fixed in the new units, then the distance between the galaxies
must shrink by a corresponding factor (recall that in (\ref{units})
$\alpha=0$, and $\beta\neq 0$).  Hence the fixed $c$ dual
of the VSL theories we have proposed contain wormhole like solutions
even without the presence of gravitating matter. This is due to the
fact that the gravitational action is indeed very complicated in the
dual picture (notice that the required unit transformation is a 
combination of VSL and conformal transformations).

Elsewhere \cite{haav} we shall show how fast-tracks  may appear in other 
theories, eg in the Bekenstein changing $\alpha$ theory. In such 
theories $\alpha$ is much smaller inside the string core, but
all other couplings remain unchanged. Hence a nomadic twin
will age much slower during  the trip, since we age electromagnetically
\cite{anthr}. 
Strong interactions just provide the nuclei for all the atoms 
we are made of. But we are essentially made of stable nuclei.
Hence if all our nuclei aged a million years we would not notice
it. Naturally in such theories one cannot avoid different aging rates between 
nomadic and sedentary twins - the curse of space travel.

\section{Black Holes with an edge}  
In \cite{vcc} we shall examine vacuum spherically symmetric solutions 
to all these theories. They have a common feature which can be
illustrated by the well-known solution in Brans-Dicke theory (which
is $a=4$, $b=0$). Using the isotropic form of the metric:  
\be  
ds^2=-fdx^{0^{2}}+g
[dr^2 +r^2(d\theta^2 +\sin^2\theta d\phi^2)]  
\ee  
we have   
\bea  
f&=&f_0{\left(1-{B\over r}\over 1+{B\over r}\right)}^{2/\lambda}\\  
g&=&g_0(1+B/r)^4
{\left(1-{B\over r}\over 1+{B\over r}\right)}^{2(\lambda-C-1)/\lambda}\\
\psi&=&{-C\over 4\lambda}\log{\left(1-{B\over r}\over 1+{B\over r}\right)}
\eea  
where $f_0$, $g_0$, $B$, and $C$, are constants, with:
\be
\lambda= [(C+1)^2-C(1-\omega_{bd}C/2)]^{1/2}
\ee
Expressions for these constants in terms of the black hole mass $m$
and coupling $\omega_{bd}$ may be found in \cite{dick2}. 
As we approach the horizon ($r_h=B$) we find that 
$c$ goes to either zero or infinity (like $(r-r_h)^N$ with $N$
related to $\omega_{bd}$).
The implication is obvious: for some parameters of the theory
(in this case requiring $q\neq 0$\footnote
{In plain Brans-Dicke ($a=4,b=0,q=0$) we have that
$c\rightarrow 0$ but $\tau\rightarrow
\infty$ in such a way that particles may enter the horizon. Hence
the discussion presented here does not apply, as one would expect.}) 
no observer may 
enter the horizon. The number of cycles
of an interaction clock trying to enter the horizon is given by:
\be 
\int^{r_h} {dt\over \tau}= \int^{r_h} {dr\over v\tau}=
\int^{r_h} {dr\over c^{2q+2}}
\ee 
which diverges for $2(q+1)N>1$. 

Again this phenomenon may be interpreted variously, 
 depending on which units are used, but all interpretations  lead
to the same physical conclusion (which is dimensionless): particles
are unable  to enter the horizon. In VSL units particles
cannot enter the horizon because they stop as $c$ goes to zero. In 
fixed-$c$ units they cannot enter the horizon because the time rates
of all interactions go to zero (as all couplings go to infinite). 
Old age strikes before anything ``has time'' to enter the horizon. 

Naturally finite sized bodies suffer from further effects,
analogous to tidal stresses, since they will probe gradients in $c$. 
Since $v\propto c$ they get squashed if $c\rightarrow 0$, or get  
stretched otherwise. $c$-induced changes of pace also induce  
gradients of aging across finite-sized bodies.   

A pedagogic illustration, studied further 
in \cite{vcc}, is a muon produced close to the black hole, moving towards
its horizon.  Such a set up is useful, for instance, when trying to  
convince skeptics of the physical validity of time dilation, or Lorentz  
contraction (eg. the fate of cosmic ray muons entering the atmosphere).  
To an Earth observer,  
if time dilation was not a physical effect the muon should never hit the  
surface of the Earth. From the point of view of the muon, if the atmosphere  
did not appear Lorentz contracted, it should have decayed before hitting  
the surface. The same set up will be of assistance here. 
No matter how close to the horizon the muon is produced, it never
reaches it.  In VSL units
the muon stops as it tries to enter the horizon, because its speed
is close to $c$, but $c$ goes to zero.
In fixed-$c$ units the muon moves close to the (constant) speed of light,
but its lifetime goes to zero as it tries to enter the horizon.
From either perspective the muon can never enter the horizon.

``Horizon'' is therefore a misnomer, and we relabel it an ``edge'':
a boundary where $c$ goes to zero sufficiently fast that no
object may reach it. On physical grounds we should postulate that 
regions beyond the edge be  excised from the manifold. Then
VSL manifolds may have an edge.

We arrive at a similar conclusion to Section~\ref{empty}. 
VSL and fixed-$c$
units are locally but not globally equivalent. The VSL picture may be
globally  more clear (in the case $a=4$, only if $q\neq 0$). 
It builds into space-time the topology perceived by actual physical
processes, in this case  excising regions which 
are physically inaccessible.

The implications for the theory of singularities are quite impressive. 
Even though we have a singularity at $r=0$, 
it is physically inaccessible. One may be able to prove that
all singularities are subject to the same constraint. 
This situation was discussed in \cite{quir}. It looks as if a stronger version
of the cosmic censorship principle might apply to these theories. 
  
\section{Conclusion}  
One must sympathise with the view that VSL theories are   
rendered objectionable by their outright violation of Lorentz invariance.  
However, previous attempts to make the Albrecht-Magueijo model  
``geometrically honest'' 
were no  
less ugly than the original; and were useless for cosmology. 
In this paper we proposed a geometrically honest VSL theory,  
corresponding to a theory in which all fine structure constants 
are promoted to dynamical variables. A changing charge interpretation 
in unnecessarily complicated - so we adopt units in which $c$ changes, 
leading to a simple picture. This should not scandalise anyone. 
 
All these theories are locally Lorentz invariant, and covariant
in a sense incorporated by a generalized structure.  
We find that physics lives on a fibre bundle. Usually physics takes 
place on the tangent bundle. At each point in space-time there is 
a tangent space, corresponding to free falling frames in which physics 
is Minkowskian. We have a similar construction, but in the new units 
the space is not the tangent space of any coordinate patch in the manifold. 
It is still a vector space - but it is not the tangent vector space, 
except in the rare cases where the change of units is holonomic
\footnote{ 
The situation is more complicated for a gravitation theory based 
upon Fock-Lorentz space-time. Now the physics' space at a given point  
is no longer a vector space, but a projective space. The fibre bundle 
multiplies the base manifold by a projective space at each point.}.

Given that a change of units maps these structures to standard   
covariant and local Lorentz invariant theories, one may wonder why   
it is worth bothering. 
To answer this question, throughout this paper we examined
these ``dual'' theories.  
For them $c$ is a constant (as well as $\hbar$ and possibly $G$), 
but naturally other quantities must vary. Indeed all couplings 
must change, at fixed ratios.   
We therefore have a theory not dissimilar from Bekenstein's changing  
$e$ theory, but such a picture is
horribly misleading for the following reasons.
  
Firstly all charges, not only $e$, will vary. But
they  vary at constant ratios, so that all changes   
may be attributed to a change in $c$ alone. Hence the dual theory  
is a theory which promotes coupling constants to dynamical variables,  
but then only allows rather contrived variations, ie variations which  
may be absorbed into a changing $c$. It seems therefore more natural  
to consider a changing $c$ description, even though the two descriptions  
are indeed operationally equivalent.   
  
Secondly the minimal dynamics in the two frames is totally different.  
This results from the fact that the action has units, and therefore  
changes under a change of units mapping dual theories.   
The minimal Bekenstein-type of theories does not have the same coupling to  
gravity as appears in the minimal VSL formulation. Rewriting
the Lagrangian of minimal VSL theories in fixed $c$ units leads to
an unpleasant mess (Section~\ref{duals}).
  
Thirdly and more importantly, the correspondence between VSL and its
duals is only local. Globally the VSL picture can be more clear.
We gave two striking examples. Fock-Lorentz space-time is just a change
of units applied to Minkowski space-time; however it contains $t>\infty$
extensions to Minkowski space-time which are physically accessible.
The horizon of a black hole may be physically impenetrable, since
$c$ goes to zero. Calling it an edge, and excising the bit beyond the 
edge seems reasonable. In the dual picture no warning about the
fact that a piece of the manifold is inaccessible is given. It is
an afterthought to notice that all interaction strengths force
the pace of aging to become very fast; thereby, for all practical  
purposes, preventing anything from entering the horizon.

Hence the VSL theories we have proposed are changing $c$ theories
simply because choosing units in which $c$ varies leads to a simpler
description. Their underlying geometrical structure is that of 
standard fixed $c$ theories subject to a change
of units; a fact undeniably placing them 
at the pinnacle of geometrical honesty.
It remains to show that
these theories, applied to cosmology, perform as well as the
Albrecht and Magueijo model. Such is the purpose of 
\cite{vcb}.
In any case it is not difficult to guess the overall cosmological 
picture to emerge in 
these theories. We see that the presence of a cosmological 
constant $\Lambda$ generally drives changes in $c$, 
which in turn convert the vacuum energy into radiation leading to a 
conventional Big Bang. However, such a Big Bang is free from the
standard cosmological problems, including the cosmological constant
problem. The fact that particle production occurs naturally
in these theories
ensures that we also solve the entropy problem. 

\section*{Acknowledgements}I would like to thank Andy Albrecht, John
Barrow, Kim Baskerville, Carlo Contaldi, Tom Kibble, and Kelly Stelle
for help in connection with this paper. I am grateful to 
the Isaac Newton Institute for support and hospitality while
part of this work was done.

\section*{Appendix - Bimetric realization of the Albrecht-Magueijo model}  
A theory which emulates many of the features of the Albrecht and Magueijo  
model (except for breaking Lorentz invariance) is the following.  
Let there be two metrics, $g$ coupling to gravitation and matter, and  
$h$ coupling to the field $c$ only. Then we may take the following  
action:  
\bea\label{S3}    
S&=& S_1+S_2\nonumber\\ 
S_1&=& 
\int d^4x \sqrt{-g}{\left(  
R-2\Lambda +{ 16\pi G\over c_0^4e^{4\psi}}{\cal L}_m \right)} \nonumber\\ 
S_2&=& 
\int d^4 x\sqrt{-h}{\left( H-2\Lambda_h 
-\kappa h^{\mu\nu}\partial_\mu\psi\partial_\nu\psi  \right)} 
\eea  
where $g_{\mu\nu}$ and $h_{\mu\nu}$ lead to two Einstein tensors 
$G_{\mu\nu}$ and $H_{\mu\nu}$, and $\Lambda$ and $\Lambda_h$ are 
their respective (geometrical) cosmological cosntants.  
Varying with respect to $g$, $\psi$, and $h$ leads to equations  
of motion:  
\bea  
G_{\mu\nu}+\Lambda g_{\mu\nu}&=&{8\pi G\over c_0^4e^{4\psi}}    
T_{\mu\nu}\\  
\Box_h \psi&=&{32\pi G\over c_0^4e^{4\psi}\kappa}\sqrt{g\over h}  
{\cal L}_m\\  
H_{\mu\nu}+\Lambda_h h_{\mu\nu}&= & -{\kappa\over 2}    
{\left(\nabla_\mu\psi \nabla_\nu\psi    
-{1\over 2}h_{\mu\nu}\nabla_\alpha\psi \nabla^\alpha\psi    
\right)}    
\eea  
Hence we may derive from an action principle the property that the  
field $\psi$ does not contribute to the stress-energy tensor  
which acts as a source to normal space-time curvature.   
In \cite{vcb,vcc} we shall highlight some curiosities 
pertaining to these theories.

\end{document}